\documentclass[aps,prl,prb,twocolumn,showpacs,superscriptaddress,groupedaddress]{revtex4-2}
\usepackage{graphicx}
\usepackage{physics}
\usepackage{amssymb}
\usepackage{amsmath}
\usepackage{latexsym}
\usepackage{ulem}
\usepackage{xcolor}
\usepackage{dcolumn}
\usepackage{subfigure}
\usepackage[T1]{fontenc}
\usepackage[utf8]{inputenc}
\usepackage[english,french]{babel}
\usepackage[colorlinks=true,linkcolor=blue,citecolor=blue]{hyperref} %
\usepackage{bm}        
\usepackage{amssymb}   

\usepackage[capitalize]{cleveref}

\usepackage{lipsum}

\begin{document}

\graphicspath{{figures/}}

\definecolor{airforceblue}{rgb}{0.36, 0.54, 0.66}

\title{Constrain relations for superfluid weight and pairings \\ in a chiral flat band superconductor}

\author{M. Thumin}%
 \email{maxime.thumin@neel.cnrs.fr}
 \author{G. Bouzerar}%
 \email{georges.bouzerar@neel.cnrs.fr}
\affiliation{Université Grenoble Alpes, CNRS, Institut NEEL, F-38042 Grenoble, France}%
\date{\today}

\begin{abstract}
\textbf{Abstract} – Within ten years, flat band (FB) superconductivity has gained a huge interest for its remarkable features and connection to quantum geometry. We investigate the superconductivity in a FB system whose orbitals are inequivalent and in which the gap and the quantum metric are tunable. 
The key feature of the present theoretical study is to show a unique and simple constrain relation that pairings obey. 
Furthermore, pairings and superfluid weight in partially filled FB are shown to be controlled by those of the half-filled lattice.
We argue that the geometry of the lattice or the complexity of the hopping terms have no impact on the features revealed in this work as far as the system is bipartite.
\end{abstract} 

\maketitle


\textbf{Introduction.} – Graphene based materials have popularized a new kind of unconventional superconductivity (SC) \cite{Cao2018_1,Cao2018_2, Efetov,HOPG_2000, HOPG_2012}, taking its origin in narrow bands. In the limit of vanishing bandwidth, electrons acquire infinite effective mass and $k$-independent dispersion relation called flat bands (FBs). These electronic states can be localized by means of quantum destructive interference \cite{Sutherland_CLS}, yielding an insulating ground state. Because of the quenching of the kinetic energy, the electron-electron interaction gives rise to a wide range of quantum phases \cite{Synthese_FB, Lieb_Lattice_Origine,Tasaki_magnetism,Frac_Hall_1, Frac_Hall_2, Frac_Hall_3}, and similarly with bosonic FBs in optical lattices \cite{Wigner_FB, BEC_FB}.
In particular, attractive interactions in FBs lead to bound-state formation with finite effective mass by a Cooper-like process \cite{Cooper_pair_peotta, Iskin_effective_mass, Torma_revisiting}, and transport of superfluid (SF) nature arises \cite{Peotta_performed_pairs}. The FB induced SC had been shown to be quantitatively different from that of single band BCS physics. The order parameter $\Delta$ and the critical temperature $T_c$ scale linearly with the interaction strength $|U|$ \cite{Gap_linear_1990, Gap_linear_1994, Volovik_T_linear, Volovik_T_linear_Flat_bands_in_topological_media} while BCS theory predicts $\Delta \propto k_BT_c \propto t \, e^{-1/\rho(E_F)|U|}$ \cite{BCS}. These unusual features in FBs have a geometrical interpretation, and can be connected to the concept of quantum metric (QM) \cite{Peotta_Nature} as defined in Ref.\cite{Provost_metric}. \\
In recent studies, assuming a uniform local pair formation $\Delta_\Lambda$ in each sublattice ($\Lambda = \mathcal{A}$ or $\mathcal{B}$), it has been shown in the case of half-filled bipartite lattices that: $N_\mathcal{B}\Delta_\mathcal{B}-N_\mathcal{A}\Delta_\mathcal{A}=\frac{|U|}{2}(N_\mathcal{B}-N_\mathcal{A})$, $N_\Lambda$ being the number of orbitals on the sublattice $\Lambda$ \cite{Peotta_Lieb, Torma_revisiting}. \\
Within the Harper-Hubbard model and under the UP hypothesis, it has been shown that the pairings and the SF weight in partially filled FBs are given by their values at half-filling, reduced by a coefficient which depends on the filling factor only \cite{Peotta_Nature}. 
However, in realistic multi-band or complex systems, orbitals are often inequivalent and thus pairings as well.
Hence, a natural question rises: Does these relations hold in such systems? 
Our main goal, is to address this issue. Furthermore, here, we shed light on a novel hidden relation between the pairings in multi-band systems which to the best of our knowledge has not been revealed so far. It concerns the case where the electron densities correspond to chemical potential inside the dispersive bands (DBs).
To answer these open questions, we consider the simplest one-dimensional bipartite FB lattice: the  stub lattice (SL). It possesses three inequivalent orbitals, a tunable gap and a tunable QM as well. Despite the absence of phase transition at finite $T$ \cite{Mermin_Wagner, Hohenberg}, one and quasi-one-dimensional lattices are systems that facilitate the understanding of the underlying physics, and relevant as well for the description of SC in anisotropic systems \cite{Quasi_1D_Cs2Cr3As3, Quasi_1D_Nature, Quasi_1D_NaMn6Bi5}. \\
We choose to tackle the FB superconductivity in the SL, within the mean-field Bogoliubov-de-Gennes (BdG) approach.
At first glance, this choice may appear questionable. Indeed, it is well known that in one dimensional systems, the quantum and thermal fluctuations have their most dramatical effects. 
However, it has been shown over the last decade that BdG is not only a qualitative tool but has proven to be a reliable method to study quantitatively the physics in the presence of FBs, even in low dimensional systems. For instance, in the sawtooth chain and in the Creutz ladder, the calculation of superfluid weight $D_s$ by BdG and Density Matrix Renormalization Group (DMRG) has revealed an impressive quantitative agreement \cite{Batrouni_sawtooth, Batrouni_Designer_Flat_Bands}.
The localized character of the FB eigenstates (or CLS) is at the origin of this astonishing ability to capture accurately the many body physics. \\

\begin{figure}[h!]
\centering
\includegraphics[scale=0.4]{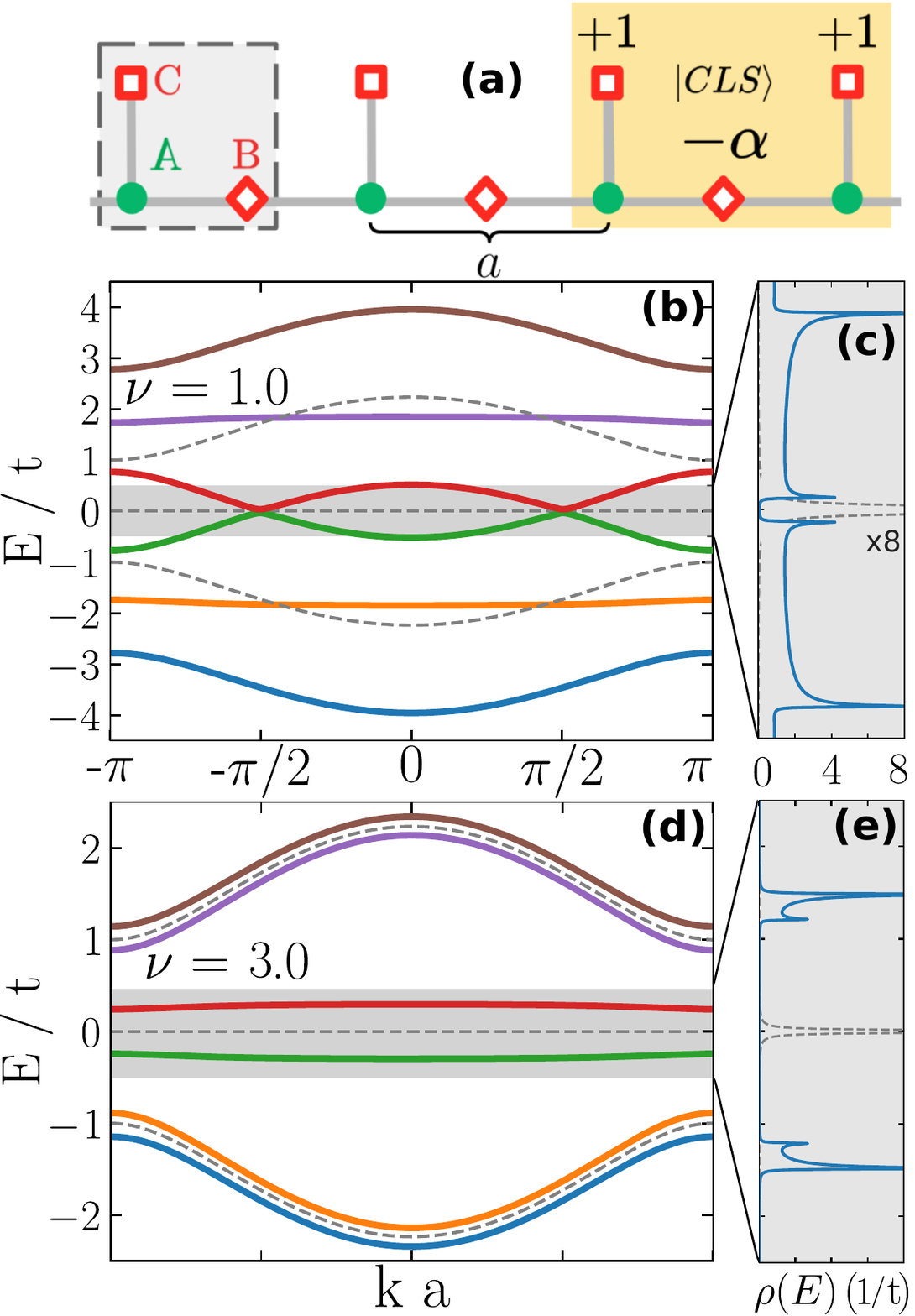} 
\caption{\textbf{(a)} Representation of the SL and a compact localized FB state (CLS). \textbf{(b)} and \textbf{(d)}: QP dispersions for $|U|/t=1$ and $\alpha=1$, with $\nu=1$ (respectively $\nu=3$).  
\textbf{(c)} and \textbf{(e)}: QP density of states for $E/t \in [-0.5,0.5]$ corresponding to the shaded areas in \textbf{(b)} and \textbf{(d)}. 
The grey dashed lines pictures the two-folded degenerated single-particle dispersions ($U=0$).}
\label{Fig. 1}
\end{figure}

\textbf{Model and methods.} –
The SL is a one-dimensional bipartite lattice with three orbitals ($A,B$, and $C$) per cell as illustrated in Fig$\,$\ref{Fig. 1}(a). Such lattice has one orbital (A) on sublattice $\mathcal{A}$ and two (B,C) on sublattice $\mathcal{B}$. This implies, from the chiral symmetry, the existence of a FB regardless the values of the $\mathcal{A}-\mathcal{B}$ hoppings. This lattice has no natural realization, but it can be engineered artificially with optical lattices \cite{stub_photonics}, or micro-pillar optical cavities \cite{stub_optical_cavities}. \\
Electrons are described by the attractive-Hubbard model that reads,
\begin{equation}
\begin{split}
    \hat{H} = \sum_{\langle i\lambda,j\eta \rangle, \sigma} t^{\lambda\eta}_{ij} \; \hat{c}_{i\lambda, \sigma}^{\dagger} \hat{c}_{j \eta, \sigma} - \mu \hat{N} - |U| \sum_{i\lambda} \hat{n}_{i\lambda\uparrow}\hat{n}_{i\lambda\downarrow}
    \label{H_exact}
\end{split}
,
\end{equation}
where $\hat{c}_{i \lambda \sigma}^{\dagger}$ creates an electron of spin $\sigma$ at site $\textbf{r}_{i\lambda}$, $i$ being the cell index and $\lambda=A,B$ and $C$ the orbitals. Sums run over the lattice, $\langle i\lambda,j \eta\rangle$ refers to nearest-neighbor pairs for which the hopping integral $t^{\lambda\eta}_{ij}$ is $t$ for $(A,B)$ pairs and $\alpha t$ for $(A,C)$ pairs. The particle number operator is $\hat{N}=\sum_{i\lambda,\sigma} \hat{n}_{i\lambda,\sigma}$, and $\mu$ is the chemical potential. Finally, $|U|$ is the strength of the on-site attractive electron-electron interaction. Notice that $\alpha$ allows as well the tuning of the QM.
\\
The BdG approach consists in decoupling the Hubbard interaction term as follows,
\begin{equation}
\begin{split}
\hat{n}_{i\lambda,\uparrow}\hat{n}_{i\lambda,\downarrow} \overset{BdG}{\simeq}
&\langle\hat{n}_{i\lambda,\downarrow}\rangle _{th} \hat{n}_{i\lambda,\uparrow} +\langle\hat{n}_{i\lambda,\uparrow}\rangle _{th} \hat{n}_{i\lambda,\downarrow} \\
+ \; &\,\frac{\Delta_{i\lambda}}{|U|} \hat{c}^{\dagger}_{i\lambda,\uparrow}\hat{c}^{\dagger}_{i\lambda,\downarrow} 
+ \frac{\Delta_{i\lambda}^*}{|U|} \hat{c}_{i\lambda,\downarrow}\hat{c}_{i\lambda,\uparrow}  \\ 
- \Big[ &\langle\hat{n}_{i\lambda,\uparrow}\rangle _{th} \langle \hat{n}_{i\lambda,\downarrow} \rangle _{th} + \Big|\frac{\Delta_{i\lambda}}{U}\Big|^2 \; \Big] , 
\end{split}
\end{equation}
where $\langle \ldots \rangle _{th}$ is the grand canonical average value. $-|U| \hat{n}_{i\lambda,\sigma} \langle\hat{n}_{i\lambda,-\sigma}\rangle _{th}$ is the Hartree term (self-consistent on-site potential), and $\Delta_{i\lambda} = -|U|\langle\hat{c}_{i\lambda,\downarrow}\hat{c}_{i\lambda,\uparrow}\rangle _{th}$ is the pairing that measures the formation of Cooper pairs. Translation invariance implies that the average $\langle\ldots\rangle _{th}$ of a local operator is cell-independent, thus, we drop the cell index. 
For a fixed temperature and a given chemical potential, pairings and occupations are calculated self-consistently. Here, we set $T=0$, and smoothly change the chemical potential to cover all possible fillings $\nu$, which corresponds to the number of electron per unit cell. We assume as well a non-magnetic ground-state,  $\langle\hat{n}_{\lambda,\uparrow}\rangle_{th}=\langle\hat{n}_{\lambda,\downarrow}\rangle_{th}= n_{\lambda}/2$. 
Within this formulation, the Hamiltonian \eqref{H_exact} becomes,
\begin{equation}
    \hat{H}_{BdG} = \sum_k 
    \begin{bmatrix}
        \hat{c}_{k \uparrow}^{\dagger} & \hat{c}_{-k \downarrow} \\ 
        \end{bmatrix} 
       \begin{bmatrix} 
         \hat{h}_{\uparrow}(k)& \hat{\Delta} \\ 
         \hat{\Delta}^{\dagger}& -\hat{h}_{\downarrow}^*(-k) \\
   \end{bmatrix}
          \begin{bmatrix} 
         \hat{c}_{k \uparrow} \\ 
         \hat{c}_{-k \downarrow}^{\dagger} \\
   \end{bmatrix}
   ,
   \label{BdG}
\end{equation}
where $\hat{c}_{k\sigma}^{\dagger} = \Big(\hat{c}_{k A,\sigma}^{\dagger}, \hat{c}_{kB,\sigma}^{\dagger}, \hat{c}_{kC,\sigma}^{\dagger}\Big)$, $c_{k\lambda,\sigma}^{\dagger} $ is the Fourier transform (FT) of $c_{i\lambda,\sigma}^{\dagger}$.  $\hat{h}_\sigma (k)=\hat{h}_0(k)-\mu \hat{\mathbb{I}}-\hat{V}_{\sigma}$ where $\hat{h}_0$ is the FT of the tight-binding term in Eq.$\,$\eqref{H_exact}, $\hat{V}_\sigma=\frac{|U|}{2}\,\text{diag}(n_{A},n_{B},n_{C})$ and $\hat{\Delta}=\text{diag}(\Delta_A, \Delta_B, \Delta_C)$. Notice that, within the UP condition, the pairing matrix becomes $\hat{\Delta}=\text{diag}(\Delta_{\mathcal{A}}, \Delta_{\mathcal{B}}, \Delta_{\mathcal{B}})$.
\begin{figure*}
    \centering
\includegraphics[scale=0.85]{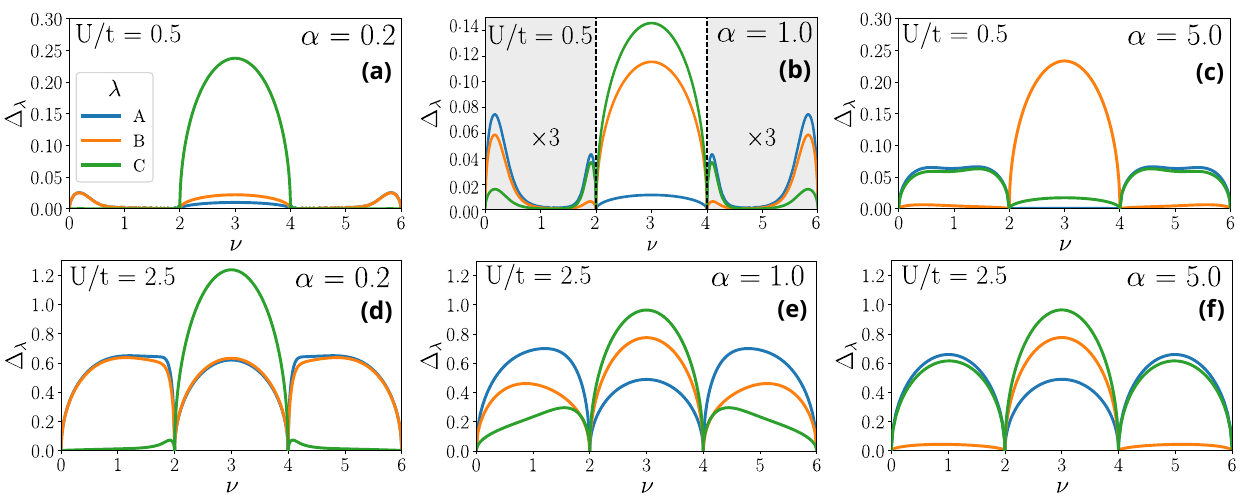}    
     \caption{\textbf{(a)} Pairings $\Delta_\lambda$ ($\lambda = A,B,C$) in unit of $t$, as a function of the filling factor $\nu$. \textbf{(a)-(c)} correspond to $|U|/t=0.5$ and respectively $\alpha = 0.2, 1.0$ and $5.0$ and \textbf{(d)-(f)} correspond to $|U|/t=2.5$ and $\alpha=0.2, 1.0$ and $5.0$. In \textbf{(b)}, $\Delta_\lambda$ in DB regions (see main text) were multiplied by a factor of three.}
 \label{Fig. 2}
\end{figure*}
\\ 

Before discussing our results, we briefly recall the procedure to access partial fillings of the FB. The eigenstates being degenerate, this band can not be partially filled by tuning the chemical potential. However, one can obtain the solution for the half-filled case corresponding to $\mu=-\frac{|U|}{2}$, and $n_\lambda=1$ according to the uniform density theorem \cite{Th_Lieb_uniform}. Pairings are obtained self-consistently. For this specific value of $\mu$, the Hubbard Hamiltonian possess a remarkable pseudo-spin SU(2) symmetry \cite{Yang_SU(2)_sym, Zhang_SU(2)_sym}. The expectation value of the pseudo-spin is defined as,
\begin{equation}
\langle \hat{\textbf{T}}_\lambda \rangle_{th} =
\begin{bmatrix} 
          - \eta_\lambda \Re \Delta_\lambda / |U| \\ 
          - \eta_\lambda \Im \Delta_\lambda / |U| \\
          ( n_{\lambda}-1 ) / 2 \\ 
\end{bmatrix}
,
\end{equation}
where the variable $\eta_\lambda$ is $-1$ on sublattice $\mathcal{A}$ and $+1$ on sublattice $\mathcal{B}$.
Any rotation of the pseudo-spins leaves the BdG Hamiltonian invariant, and allows to access partially filled FB solutions. Notice that this procedure has been used in Refs \cite{Batrouni_CuO2, Peotta_Lieb} A detailed presentation of the pseudo-spin SU(2) symmetry can be found in the Supplemental Material \cite{SM}. \\

\textbf{Results and discussions.} –
In Fig.$\,$\ref{Fig. 1}(b) and (d), the quasi particle (QP) dispersions are plotted for $\alpha=1$, and for $|U|/t=0$ (dashed lines) and $|U|/t=1$ (full lines). 
The carrier density is $\nu=1.0$ in Fig.\ref{Fig. 1}$\,$(b) and $\nu=\Bar{\nu}=3.0$ (half-filling) in Fig.\ref{Fig. 1}$\,$(d). Because of the SU(2) pseudo-spin symmetry, the dispersion is unchanged for any $\nu \in [2,4]$.
First, for $|U|=0$, a gap $\delta_0=|\alpha|t$ opens up in the one particle spectrum between the FB and the DBs at $k=\pi$. 
When $|U|$ is switched on, the two-fold degeneracy of the bands is lifted. 
For $\nu=1.0$, corresponding to the half-filled lower single-particle band, a QP gap opens up at $k=\frac{\pi}{2}$ which can be seen as well in the density of states as well.
The lowest QP excitation is well described (as expected) by the standard BCS QP-dispersion \cite{BCS}, but because of the Hartree terms, the chemical potential should be renormalized. Unsurprisingly, the QP gap grows as $\Delta_{QP} \propto t \, e^{-A/|U|}$. 
On the other hand, as discussed in the introduction, when $\nu$ corresponds to partially filled FB ($\nu \in [2,4]$), the SC is inconsistent with conventional BCS theory, now the QP gap scales linearly with $|U|$. Additionally, because of the asymmetry between $B$ and $C$ orbitals, the FBs acquire a finite bandwidth which scales as well linearly with $|U|$. In contrast, in the Lieb lattice, these QP bands remain dispersionless. \\

Figure$\,$\ref{Fig. 2} depicts the pairings $\Delta_{\lambda} = -|U|\langle\hat{c}_{i\lambda,\downarrow}\hat{c}_{i\lambda,\uparrow}\rangle _{th}$ as a function of $\nu$. Subplots (a),(b) and (c), correspond to $|U|/t=0.5$ and respectively $\alpha=0.2, 1.0$ and $5.0$, while (d), (e), (f) correspond to $|U|/t=2.5$ and the same values of $\alpha$. Note that the pairings are taken real since they all have the same phase which can be removed by global gauge transformation.
Due to the electron-hole symmetry, properties are identical for the densities $\nu$ and $6-\nu$. It is convenient for what follows to define three different regions: LDB region for $\nu \in [0,2]$, UDB region for $\nu \in [4,6]$, and FB region for $\nu \in [2,4]$. \\
Let us start with qualitative remarks to understand the role of $|U|$ and $\alpha$. First, as $\alpha$ increases at fixed $|U|$ (see (a,b,c) or (d,e,f)), the pairings in each region tend to acquire a parabolic-like shape. This is due to the increasing single particle gap $\delta_0=|\alpha| t$ which isolate bands from each others.
At fixed $\alpha$, as $|U|/t$ is tuned, one observes a double peak structure in the LDB region (respectively UDB region) where the pairings are highly asymmetric. However, as $|U|$ increases further (strong coupling regime), these peaks merge into a single one.
Notice that pairings in the FB region remain parabolic-like with a maximum at half-filling, and vanishes at $\nu=2$ (respectively $\nu=4$).
More quantitatively, we remark that the amplitude of $\Delta_\lambda$ is always larger in the FB region than in the LDB and UDB regions due to the exponential suppression of the pairings in DBs as compared to the linear scaling in the FB. 
Secondly, while $\Delta_A$ is always the largest in the DB regions, and the smallest in the FB region where $\Delta_B$ or $\Delta_C$ naturally dominates. \\

As discussed in the introduction, it has been shown in Ref. \cite{Torma_revisiting} that the pairings in the FB region obey $N_\mathcal{B}\Delta_\mathcal{B}-N_\mathcal{A}\Delta_\mathcal{A}=\frac{|U|}{2}(N_\mathcal{B}-N_\mathcal{A})$. This equation has been derived assuming a UP on each sublattice, which does not, in general, correspond to the true self-consistent solution of the BdG equations. Notice that this condition is fulfilled in the Lieb lattice. However, in the SL where pairings are inhomogeneous, for any $|U|$ and any $\alpha$ we find that, 
\begin{equation}
-\Delta_A + \Delta_{B}+\Delta_{C} = \frac{|U|}{2}    ,
\label{sum_rule_hf}
\end{equation}
holds exactly, in the half-filled chain.  \\
Notice that in a previous work \cite{stub_supra}, it has been shown that $\Delta_{B}+\Delta_{C}$ is $\frac{|U|}{2}$ in the weak coupling regime, but breaks down in the intermediate and large $|U|$ coupling. Equation$\,$\eqref{sum_rule_hf} explains clearly the origin of this deviation. 
Using the pseudo-spin SU(2) symmetry of the BdG Hamiltonian and the sum-rule given in Eq.$\,$\eqref{sum_rule_hf}, we find, 
\begin{equation}
    \Delta_\lambda(\nu) = f(\nu) \Delta_\lambda(\Bar{\nu}) , 
\label{rotation_Delta}
\end{equation} 
where $f(\nu)=\sqrt{(\nu-2)(4-\nu)}$. The details are given in Supplementary Material \cite{SM}.

In Fig.$\,$\ref{Fig. 3}(a) the ratio $\Delta_\lambda/\Delta_\lambda(\Bar{\nu})$ is plotted as a function of $\nu$ for different values of $|U|/t$ and $\alpha$. It is clearly seen that Eq.$\,$\eqref{rotation_Delta} is verified with high accuracy. Notice that similar results have been obtained within the UP condition \cite{Peotta_Nature}. Using Eq.$\,$\eqref{sum_rule_hf} and Eq.$\,$\eqref{rotation_Delta}, one gets $-\Delta_A+\Delta_B+\Delta_C = \frac{|U|}{2} f(\nu)$. \\
\begin{figure}[h!]
    \centering
    \includegraphics[scale=0.37]{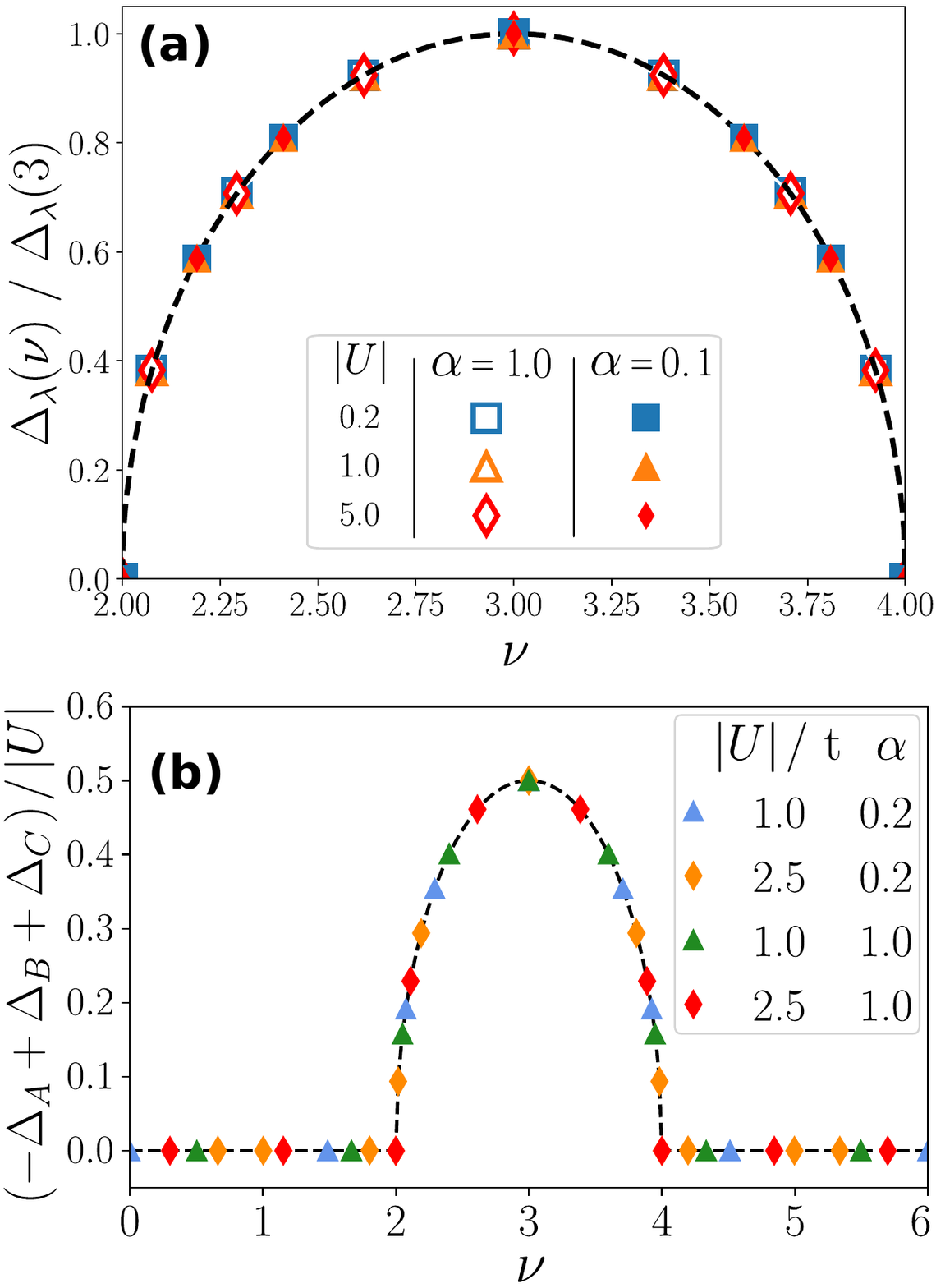} 
    \caption{\textbf{(a)} Pairings, rescaled by their value at half-filling ($\nu=3$), as a function of $\nu$ for different values of $|U|$. Empty (respectively full) symbols correspond to $\alpha=1$ (respectively $\alpha=0.1$). Black dashed line corresponds to Eq.$\,$\eqref{rotation_Delta}.  
    \textbf{(b)}$(-\Delta_A+\Delta_B+\Delta_C)/|U|$ as a function of $\nu$ for $\alpha = 0.2$ and $1.0$, and $|U|/t=1.0$ (triangles) and $|U|/t=2.5$ (diamonds).
    The black dashed line corresponds to the analytical expression $\frac{1}{2} f(\nu) \chi_{[2,4]}$ (see main text). }
    \label{Fig. 3}
\end{figure} \\
Interestingly, a careful data analysis in DB regions reveals that the pairings obey as well an additional sum-rule which reads,
\begin{equation}
    -\Delta_A+\Delta_B+\Delta_C = 0 . 
\label{sum_rule_DB}
\end{equation}
This sum-rule, or constrain, is valid for any values of $|U|$, $\alpha$ and $\mu \ne -|U|/2$. We should emphasize, that to the best of our knowledge, such a relation has never been reported in the literature. 
Finally, all these remarkable results can be included in a single equation,
\begin{equation}
-\Delta_A+\Delta_B+\Delta_C = \frac{1}{2} |U| f(\nu) \chi_{[2,4]},
\label{regle_somme} 
\end{equation} 
where $\chi_{[2,4]}$ is the indicator function of the set $[2,4]$. Equation$\,$\eqref{regle_somme} is illustrated in Fig.$\,$\ref{Fig. 3}(b). \\
It is interesting to mention that Eq.$\,$\eqref{sum_rule_hf} and Eq.$\,$\eqref{regle_somme} are numerically realized with different degree of accuracy.
Indeed, high accuracy (machine precision) is obtained at any step of the self-consistent procedure for Eq.$\,$\eqref{sum_rule_hf} while in the case of Eq.$\,$\eqref{regle_somme} it depends exclusively on the precision requested in the self-consistent loop.
\\
Additionally, we have checked that Eq.$\,$\eqref{regle_somme} is valid for other geometries, thus we argue that our findings are general and hold in any bipartite lattices. A general and rigorous proof of this statement is currently under investigation.  \\

\begin{figure}[h!]
    \centering
    \includegraphics[scale=0.35]{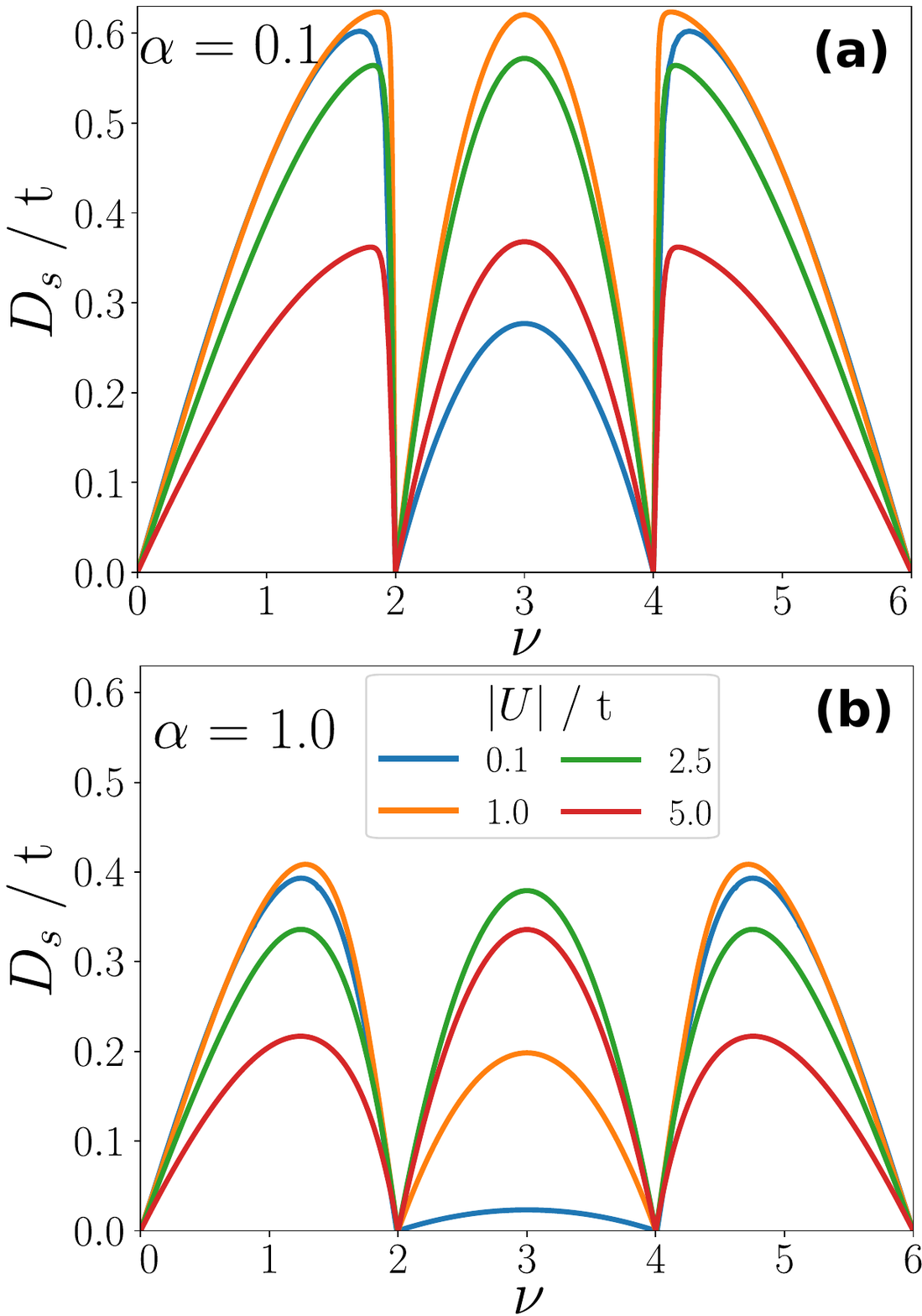} 
    \caption{\textbf{(a)} Superfluid weight $D_s$ as a function of $\nu$ for $|U|/t=0.1,1.0,2.5$. 
    In \textbf{(a)}, $\alpha=0.1$, and in \textbf{(b)} $\alpha=1.0$.}
    \label{Fig. 4}
\end{figure}

\begin{figure}[h!]
    \centering
    \includegraphics[scale=0.35]{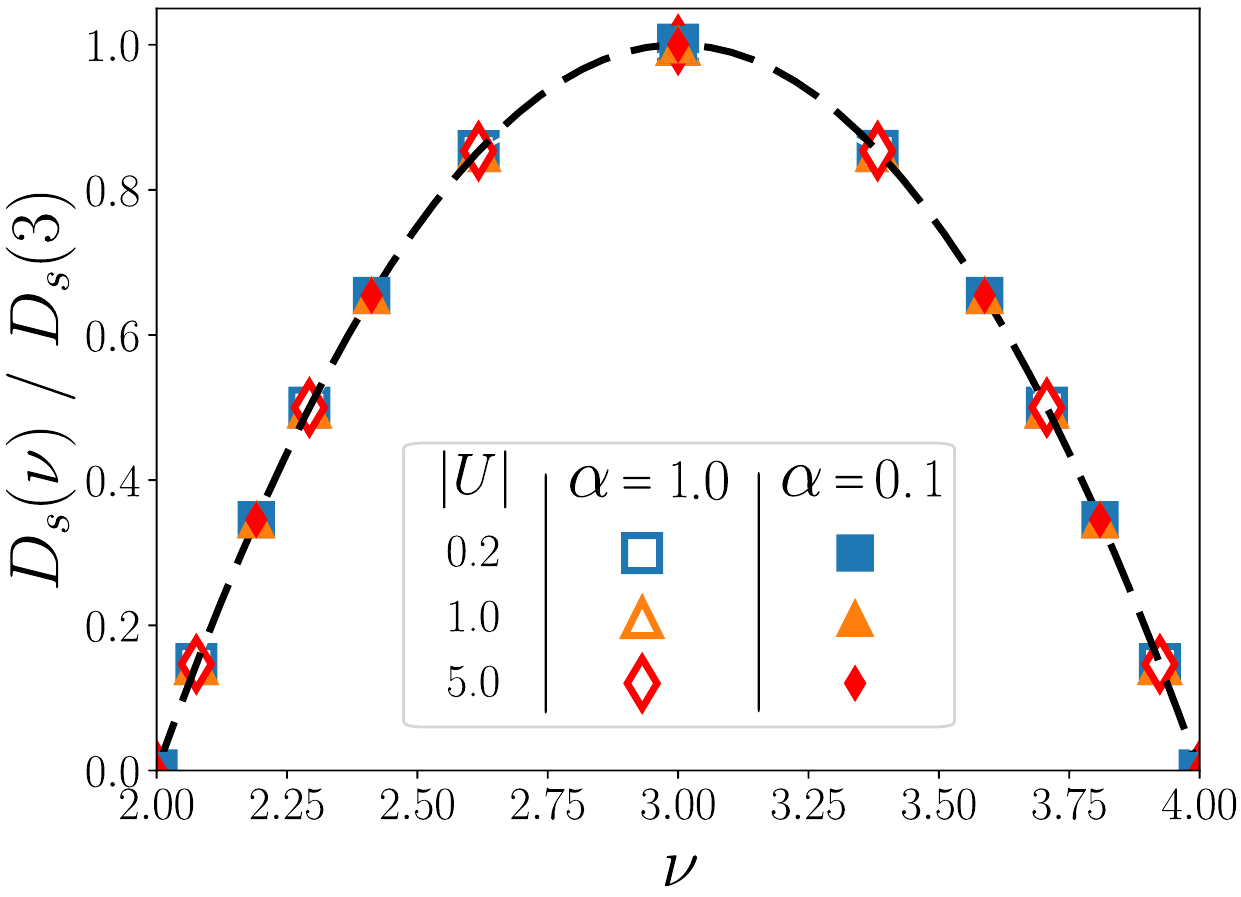} 
    \caption{Superfluid weight, rescaled by its value at half-filling ($\nu=3$), as a function of $\nu$ for different values of $|U|$. Empty (respectively full) symbols represent $\alpha=1.0$ (resp $\alpha=0.1$). Black dashed line corresponds to Eq.$\,$\eqref{rotation_Ds}. }
    \label{Fig. 5}
\end{figure}

The superconducting state is characterized by the Meissner-Ochsenfeld effect and dissipationless currents. The key quantity that
characterizes the SC phase is the SF weight \cite{Scalapino_Ds, Scalapino_criteria, Khon_Ds, Shastry_Sutherland_Ds},
\begin{equation}
D_s = \frac{1}{N_c} \frac{\partial^2 \Omega}{\partial q^2}  \Big|_{q=0} , 
\label{Ds}
\end{equation}
where $N_c$ is the number of unit cells of the lattice, $\Omega (q)$ is the grand-potential and $q$ mimics the effect of a vector potential, introduced by a standard Peierls substitution $t^{\lambda\eta}_{ij} \longrightarrow t^{\lambda\eta}_{ij} e^{iq(x_{i\lambda}-x_{j\eta})}$. \\
Figure\ref{Fig. 4}(a) pictures $D_s$ as a function of $\nu$ for $\alpha=0.1$ and different values of $|U|$. At low carrier density, one observes a linear behavior, consistent at weak coupling, with $D_s = \nu / m^*$, where $m^*=\sqrt{4+\alpha^2}/a^2$ is the effective mass at $U=0$. 
As $|U|$ increases up to $|U|/t=1$, the maximum slowly rises up to $0.62$ while its position moves towards $\nu=2$. Beyond its maximum, $D_s$ falls down rapidly to $0$. In the FB region, $D_s$  (i) has a dome-like shape with a maximum at $\nu=\Bar{\nu}$, and (ii) increases quickly as $|U|$ is switched on until it reaches a maximum for $|U|/t \simeq 1$.
For larger values of $\alpha$, many changes occur as can be seen in Fig.$\,$\ref{Fig. 4}(b). 
In the DB region, the steep drop observed at $\nu=2$ disappears and gives way to a more symmetrical shape for $D_s$.
Moreover, $D_s$ is smaller than that obtained for $\alpha=0.1$, which is a signature of weaker superconducting phase. \\
In Fig.$\,$\ref{Fig. 5}, the ratio $D_s(\nu)/D_s(\Bar{\nu})$ is plotted as a function of $\nu$ for $\alpha=0.1, 1.0$ and $|U|/t=0.2,1.0$ and $5.0$. As seen previously for the pairings plotted in Fig$\,$\ref{Fig. 3}(a), the ratio depends on $\nu$ only, regardless the values of $\alpha$ and $|U|$. 
Under the UP condition, it has been shown analytically that the $\nu$-dependent factor is $f(\nu)^2$ \cite{Peotta_Nature}. We find that it is as well the case in the SL. Because of the non-uniform values of the pairings, here the analytical proof is cumbersome. However, in both weak and strong couplings an analytical proof could be manageable.
Thus, it has been proved numerically that,
\begin{equation}
 D_s(\nu)= f(\nu)^2 D_s(\Bar{\nu}) .
\label{rotation_Ds}
\end{equation}
As discussed previously for the pairings, because of the pseudo-spin SU(2) symmetry, the SF weight at any partial filling of the FB is directly obtained
from the half-filling solution without additional calculations.\\
Before concluding, we would like to draw the reader's attention to two relevant points.
First, BdG theory and DMRG calculations are found to agree only if the Hartree terms are included in the decoupling procedure \cite{Batrouni_Designer_Flat_Bands}.
It therefore seemed useful to us to present in the Supplemental Material the impacts of the absence of Hartree terms in the BdG decoupling \cite{SM}. Secondly,
It is important to emphasize that corrections should be included in Eq.$\,$\eqref{Ds} to get the correct expression of the SF weight \cite{Torma_revisiting, Batrouni_sawtooth}. However, it has been shown that these corrections vanish when the QM is minimal which is in general realized when the position of the orbitals are maximally symmetric within the unit cell. By modifying the position of the orbitals, we have numerically checked that (i) the QM is indeed minimal for the geometry presented in Fig.$\,$\ref{Fig. 1}(a), and (ii) corrections to $D_s$ are vanishing. \\

\textbf{Conclusion.} – 
In this work, we have presented important results on unconventional flat band superconductivity in a system with inequivalent orbitals. Pairings are shown to obey a unique and simple relation which depends only on the strength of the interaction parameter and on the filling factor. 
Additionally, pairings and superfluid weight at any partial filling of the flat band can be straightforwardly obtained from the half-filled solution, as recently demonstrated assuming uniform pairings.
We argue that our results obtained in the case of the stub lattice are general, and should apply to any bipartite lattices regardless the complexity of the hoppings. \\

\pagebreak

\section*{SUPPLEMENTAL MATERIAL}
\section{I. SU(2) symmetry and partial fillings of the FB.} 
\numberwithin{equation}{section}
\setcounter{equation}{0}
\renewcommand{\theequation}{A.\arabic{equation}}
The purpose of this section is to show how to access partial fillings of the flat band (FB) within the mean-field Bogoliubov-de-Gennes (BdG) theory. This method applies only in bipartite lattices to FBs located at the center of the single particle spectrum. Here, we consider the case of the one-dimensional stub lattice, but the methodology can be applied to any bipartite lattice regardless the spatial dimension. \\
As discussed in the main text, we consider the attractive Hubbard model on the stub lattice,
\begin{equation}
\begin{split}
    \hat{H} = \hspace{-0.2cm} \sum_{\langle i\lambda,j\eta \rangle, \sigma} \hspace{-0.2cm} t^{\lambda\eta}_{ij} \; \hat{c}_{i\lambda, \sigma}^{\dagger} \hat{c}_{j \eta, \sigma} - \mu \hat{N} - |U| \sum_{i\lambda} \hat{n}_{i\lambda\uparrow}\hat{n}_{i\lambda\downarrow}
    \label{H_exact}
\end{split}
,
\end{equation}
where $\hat{c}_{i \lambda \sigma}^{\dagger}$ creates an electron of spin $\sigma$ at site $\textbf{r}_{i\lambda}$, $i$ being the cell index and $\lambda=A,B$ and $C$ the orbitals. The hoppings $t^{\lambda\eta}_{ij}$ are restricted to nearest-neighbor pairs $\langle i\lambda,j \eta\rangle$ only. The hopping value is $t$ for $(A,B)$ pairs and $\alpha t$ for $(A,C)$ pairs. 
Finally, $|U|$ is the strength of the on-site attractive electron-electron interaction, $\mu$ the chemical potential and $\hat{N}=\sum_{i\lambda,\sigma} \hat{n}_{i\lambda,\sigma}$. \\
Let us define the pseudo-spin operators $\hat{\textbf{T}}_{i\lambda}$ as, 
\begin{equation}
\hat{\textbf{T}}_{i\lambda} = \frac{1}{2}
\begin{bmatrix} 
          \quad \, \eta_\lambda \Big( \hat{c}^\dagger_{\lambda \uparrow}\hat{c}^\dagger_{\lambda \downarrow} + \hat{c}_{\lambda \downarrow}\hat{c}_{\lambda \uparrow} \Big) \\ 
          -i \eta_\lambda \Big(\hat{c}^\dagger_{\lambda \uparrow}\hat{c}^\dagger_{\lambda \downarrow} - \hat{c}_{\lambda \downarrow}\hat{c}_{\lambda \uparrow} \Big) \\
          \hat{n}_{\lambda \uparrow}+\hat{n}_{\lambda \downarrow}-1 \\
\end{bmatrix},
\end{equation}
where $\eta_\lambda=-1$ on sublattice $\mathcal{A}$ (A orbitals) and $\eta=+1$ on sublattice $\mathcal{B}$ (B,C orbitals). Using these operators, the Hamiltonian \eqref{H_exact} can be re-expressed as, 
\begin{equation}
\hspace{-0.1cm}
    \hat{H} =
    \hspace{-0.2cm}
    \sum_{\langle i\lambda,j\eta \rangle, \sigma} 
    \hspace{-0.2cm}
    t^{\lambda\eta}_{ij} \; \hat{c}_{i\lambda, \sigma}^{\dagger} \hat{c}_{j \eta, \sigma} - \mu' \hat{N} - \frac{2|U|}{3}\sum_{i\lambda} \hat{\textbf{T}}_{i\lambda} \cdot \hat{\textbf{T}}_{i\lambda}
        ,
\end{equation}
where $\mu' = \mu+\frac{|U|}{2}$. At half-filling, $\mu'=0$, and the Hamiltonian obey the commutation relations $[\hat{H},\hat{T}_{i\lambda}^\zeta]=0$, where $\zeta=x,y$ and $z$. This constitutes the SU(2) pseudo-spin symmetry \cite{Yang_SU(2)_sym, Zhang_SU(2)_sym}. Hence, the Hamiltonian is invariant under rotations of the pseudo-spin operators which we used to access the partial filling of FBs. Notice that this procedure has been used in Refs \cite{Batrouni_CuO2, Peotta_Lieb}. \\
The Hamiltonian \eqref{H_exact} is diagonalized within the BdG approach which corresponds to the decoupling: $\hat{\textbf{T}}_{i\lambda} \cdot \hat{\textbf{T}}_{i\lambda} \longrightarrow 2 \langle \hat{\textbf{T}}_{i\lambda} \rangle_{th} \cdot \hat{\textbf{T}}_{i\lambda} - \langle \hat{\textbf{T}}_{i\lambda} \rangle_{th}^2$ where the thermal average $\langle \hat{\textbf{T}}_{i\lambda} \rangle_{th}$ is given by,
\begin{equation}
\langle \hat{\textbf{T}}_{i\lambda} \rangle_{th} =
\begin{bmatrix} 
          - \eta_\lambda \Re \Delta_\lambda / |U| \\ 
          - \eta_\lambda \Im \Delta_\lambda / |U| \\
          ( n_{\lambda}-1 ) / 2 \\
\end{bmatrix}.
\end{equation}
$\Delta_\lambda$ and $n_{\lambda}$ are respectively the pairings and occupations. \\
Setting $\mu'=0$, the half-filled solution is given by uniform occupations $n_\lambda=1$ for $\lambda=A,B$ and $C$ \cite{Th_Lieb_uniform}, the pairings have to be calculated self-consistently. We define $\nu = \sum_{\lambda} n_\lambda$ the filling factor where at half-filling $\nu=\Bar{\nu}=3$. Furthermore, we assume time reversal symmetry and choose a gauge such that the pairings are real. Thus, in the half-filled system, the expectation values of the pseudo-spin operators are, 
\begin{equation}
\langle \hat{\textbf{T}}_{i\lambda} \rangle_{th} (\Bar{\nu}) =
\begin{bmatrix} 
          - \eta_\lambda \Delta_\lambda(\Bar{\nu}) / |U| \\ 
          0 \\
          0 \\
\end{bmatrix}.
\end{equation}
We now apply a rotation of angle $\theta$ around the $y$-axis $\mathcal{R}_y(\theta)$, namely,
\begin{equation}
\begin{split}
\langle \hat{\textbf{T}}_\lambda \rangle _{th} (\nu(\theta)) = 
&\underbrace{
\begin{bmatrix}
\cos{\theta} & 0 & -\sin{\theta}\\
0 & 1 & 0\\
\sin{\theta} & 0 & \cos{\theta}
\end{bmatrix}
}  
\langle \hat{\textbf{T}}_\lambda \rangle _{th} (\Bar{\nu}) , \\
& \qquad \quad \mathcal{R}_y(\theta)    
\end{split}
\label{R_y(theta)}
\end{equation} 
and get a set of pseudo-spins corresponding to partial filings of the FB. More specifically, the new solution of the BdG Hamiltonian
reads, 
\begin{equation}
    \left\{ 
\begin{array}{l l}
\Delta_\lambda(\theta) = \Delta_\lambda(\Bar{\nu}) \cos{\theta} \\
\; n_{\lambda}(\theta) =1  - \frac{2 \eta_\lambda \Delta_\lambda(\Bar{\nu})}{|U|} \sin{\theta}\\ 
 \end{array} 
 \right.  .
 \label{rotation_appendix}
\end{equation}
By tuning $\theta \in ]-\frac{\pi}{2},\frac{\pi}{2}]$ in Eq.$\,$\eqref{rotation_appendix}, we then cover all partial fillings of the FB corresponding to $2 \le \nu \le 4$. This procedure has been used to obtain the data plotted in Fig.$\, 2$ to Fig.$\, 5$  in the main text, and in Fig.$\,$\ref{Fig. 6} of this supplemental material. \\
Starting from the expression of $n_\lambda(\theta)$ as given in Eq.$\,$\eqref{rotation_appendix}, and by using the sum-rule for the pairings at half-filling (see main text) which is given by,
\begin{equation}
    - \Delta_A + \Delta_B + \Delta_C = \frac{|U|}{2},
\end{equation}
one can express the $\nu$ as a function of the angle $\theta$, 
\begin{equation}
    \nu(\theta) = 3 - \sin{\theta}.
\end{equation}
Inserting this expression in Eq.$\,$\eqref{rotation_appendix}, one finally gets the analytical expression of the pairings and occupation as a function of the filling factor,
\begin{equation}
    \Delta_\lambda(\nu) = \sqrt{(\nu-2)(4-\nu)} \Delta_\lambda(\Bar{\nu}) , 
    \label{demo_Delta}
\end{equation} 
\begin{equation}
    n_{\lambda}(\nu) = 1 - 2 \eta_\lambda (3-\nu) \frac{\Delta_\lambda(\Bar{\nu})}{|U|} . 
\end{equation}
In Eq.$\,$\eqref{demo_Delta}, $\sqrt{(\nu-2)(4-\nu)}$ corresponds to $f(\nu)$, as defined in the main text, and proves Eq.$\,$(6) of the article. \\

\section{II. Importance of the Hartree terms}
\numberwithin{equation}{section}
\setcounter{equation}{0}
\renewcommand{\theequation}{B.\arabic{equation}}
The aim of this section is to illustrate the role of the Hartree terms in the BdG decoupling.
More precisely, our goal is to find out how the superfluid (SF) weight is affected when these terms are absent. \\
In the BdG decoupling, the Hartree terms $-|U| \langle \hat{n}_{\lambda,\sigma} \rangle \hat{n}_{\lambda,-\sigma}$ are taken into account to ensure the pseudo-spin SU(2) symmetry of the Hamiltonian. In addition, as discussed in the previous section, it allows the access to the partial fillings of the FB. However, in the absence of the Hartree terms, the function $\nu(\mu)$ is no longer discontinuous at $\mu = -|U|/2$, and the FB can be continuously filled by tuning $\mu$ as we do for dispersive bands (DBs). 

\begin{figure}[h!]
    \centering
    \includegraphics[scale=0.35]{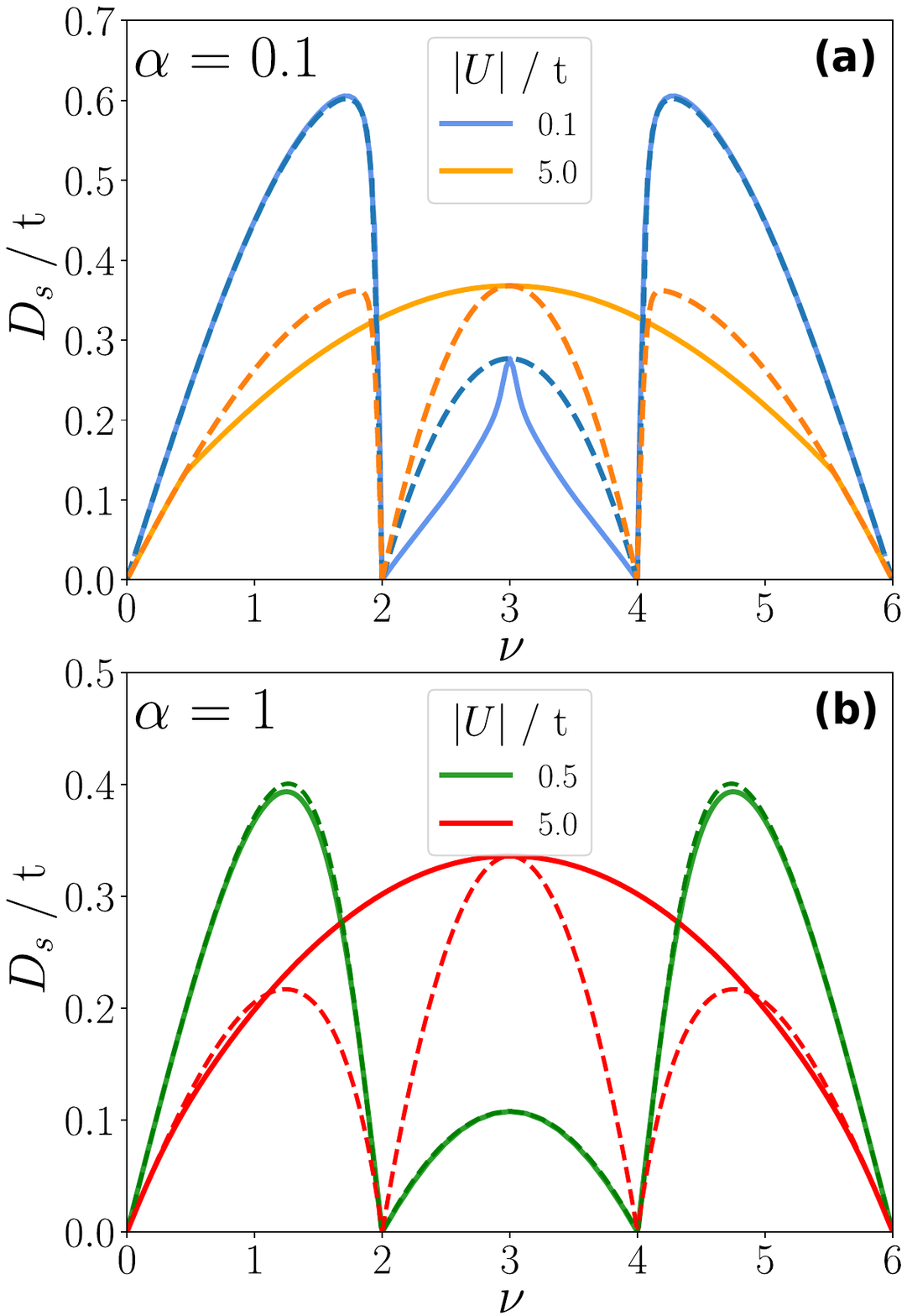} 
    \caption{SF weight as a function of $\nu$ in the weak and strong coupling regimes. In \textbf{(a)}, $|U|/t = 0.5, 5.0$ and $\alpha=1.0$. In \textbf{(b)}, $|U|/t = 0.1, 5.0$ and $\alpha=0.1$. Dashed and full lines correspond respectively to calculations with and without the Hartree terms.}
    \label{Fig. 6}
\end{figure}

Notice that in the case of a single band model, the Hartree terms can be absorbed in the chemical potential, thus they can be ignored. On the other hand, in a multi-band approach, it acts as an on-site potential or site dependent chemical potential which cannot be integrated in $\mu$. However, in the peculiar case of half-filling, the Hartree terms can be absorbed in $\mu$. \\
We now discuss, the effects of the absence of the Hartree terms on the SF weight in the case of the stub lattice . Figure$\,$\ref{Fig. 6}(a) (respectively (b)) pictures the $D_s$ as a function of $\nu$, with (dashed lines) and without (full lines) the Hartree terms, for $\alpha=1.0$ (respectively $\alpha=0.1$). \\
In Fig.$\,$\ref{Fig. 6}(a), for small $|U|$, we observe almost no change in $D_s$ when the Hartree terms are switched off, despite a minor mismatch close to the maximum in the DBs.
However, for large $|U|$, the SF weight is now drastically affected. Indeed, the three dome structure disappears and is replaced by a single dome when the Hartree terms are absent. This is the reason why in Ref.\cite{Dice_chinois}, which address the superconductivity in the dice lattice, the three domes merge into a single one as $|U|$ increases.
The presence of the Hartree terms always lead to a vanishing SF weight for both 
$\nu=2$ or $\nu=4$ (empty and fully filled FB) as predicted by Eq.$\,$(10) in the main text. \\ 
For small values of $\alpha$, as depicted in Fig.$\,$\ref{Fig. 6}(b), the conclusion is slightly different. In the weak coupling regime, there is still a good agreement in the DB regions ($\nu \in [0,2] \cup [4,6]$), but in the FB region ($\nu \in [2,4]$) the dome is now replaced by a structure which exhibits a cusp at half-filling.
Moreover, in the strong coupling regime, as seen for $\alpha=1$, a single dome is again observed. 
For filling factors in the DB regions, two regimes can be seen: (i) for $\nu \le 0.5$ Hartree terms have essentially no impact, and (ii) for $\nu \ge 0.5$, their absence leads to smaller values of $D_s$. \\
Thus, in the weak coupling regime, the Hartree terms have weak effects in the DB regions for any value of $\alpha$. On the other hand, in the FB region the Hartree terms can be ignored only when $\alpha$ is small enough. Finally, in the strong coupling regime, the absence of the Hartree terms has a dramatic impact on $D_s$ which looses its three dome structure.

\end{document}